\documentclass[11pt,twocolumn,twoside]{IEEEtran}
\usepackage{amsmath}
\usepackage[margin=0.75in,headheight=0.45in]{geometry}
\usepackage[pdftex]{epsfig}
\usepackage{amsfonts}
\usepackage{amssymb}
\usepackage{fancyhdr}
\include{graphicsx}
\usepackage{siunitx}
\usepackage[noadjust]{cite}

\usepackage{xcolor}
\newif\iffinal
\iffinal
    \newcommand\tak[1]{}
    \newcommand\ian[1]{}
    \newcommand\liz[1]{}
\else
    \newcommand\tak[1]{{\color{blue}[ tak: #1 ]}}
    \newcommand\ian[1]{{\color{red}[ Ian: #1 ]}}
    \newcommand\liz[1]{{\color{green}[ Liz: #1 ]}}
\fi

\begin{document}
\title{\vspace{0.2in}\sc Cloud Classification with Unsupervised Deep Learning}
\author{Takuya Kurihana$^{1}$\thanks{Corresponding author: I Foster, foster@uchicago.edu}\thanks{$^1$Department of Computer Science, University of Chicago}, Ian Foster$^{1,2}$\thanks{$^2$Data Science and Learning Division, Argonne National Lab}, Rebecca Willett$^{1,4}$\thanks{$^4$Department of Statistics, University of Chicago}, Sydney Jenkins$^{1,5}$\thanks{$^5$Department of Physics, University of Chicago}, \\Kathryn Koenig$^{1,6}$\thanks{$^6$Harris School of Public Policy, University of Chicago }, Ruby Werman$^{7}$\thanks{$^7$College of Letters \& Science, University of California, Berkeley}, Ricardo Barros Lourenco$^{1}$, Casper Neo$^{1}$, Elisabeth Moyer$^{3}$\thanks{$^3$Department of the Geophysical Sciences, University of Chicago}}

%Michael Maire$^{1}$
%ITF: I am adding comments prefixed by %ITF 

\maketitle
\thispagestyle{fancy}
\begin{abstract}
We present a framework for cloud characterization that leverages modern unsupervised deep learning technologies. %developed in computer vision.
%ITF: I think we want to say that a) there is a lack of labeled cloud images, preventing the use of supervised learning; and b) even if we had labeled images, the images would only correspond to human terms (cumulus, etc.) that might not be meaningful. [Would be good to have Liz's input here, as these critiques are somewhat subtle.)]
% existing classification schemes are historical and arbitrary, and we want to have them emerge form the data
%Conventional neural network-based cloud classification models have used supervised learning methods. However, an unsupervised learning method is the more appropriate and powerful approach due to a lack of labeled cloud images. In addition, unsupervised learning methods  avoid restricting the model to artificial categories based on historical classification schemes (e.g., cumulus, cirrus). %[more]
While previous neural network-based cloud classification models have used supervised learning methods, unsupervised learning allows us to avoid restricting the model to artificial categories based on historical cloud classification schemes and enables the discovery of novel, more detailed classifications.
Our framework learns cloud features directly from radiance data produced by NASA's Moderate Resolution Imaging Spectroradiometer (MODIS) satellite instrument, deriving cloud characteristics from millions of images without relying on pre-defined cloud types during the training process. 
We present preliminary results showing that our method extracts physically relevant information from radiance data and produces  meaningful cloud classes. 
\end{abstract}

\section{Motivation}
%(SCIT~\cite{scit})
Clouds play a dominant role in the Earth's radiation budget, both reflecting sunlight and trapping infrared radiation. Their responses are the principal source of uncertainty in numerical simulations of future climate, because even state-of-the-art climate models cannot accurately resolve cloud formation and evolution on scales from sub-kilometers to thousands of kilometers \cite{NICAM}. NASA satellite instruments have observed cloud behavior for several decades, providing us with a rich dataset that can potentially inform understanding of cloud dynamics and feedbacks, but these large datasets have not yet been fully employed, in part because computing power has only recently approached the necessary scale. Clouds are therefore a timely target for large scale computational analyses that can automate detection of cloud attributes and identify scientifically relevant cloud classes.
%that will leverage these large quantities of data to obtain new insights into cloud characteristics and dynamics. We may, for example be able to identify cloud classes based on features such as texture, and then observe changes in the frequency and distribution of such classes over time.  

%[One more paragraph for clouds characterization history]
%previous attempts to develop computer-enabled classifications have relied supervised learning, which cannot yield novel insights

% cloud figure 
%pacificocean_tmo_2016032_lrg.jpg
\begin{figure}[t!]
    \begin{center}
    \epsfxsize=\hsize \epsfbox{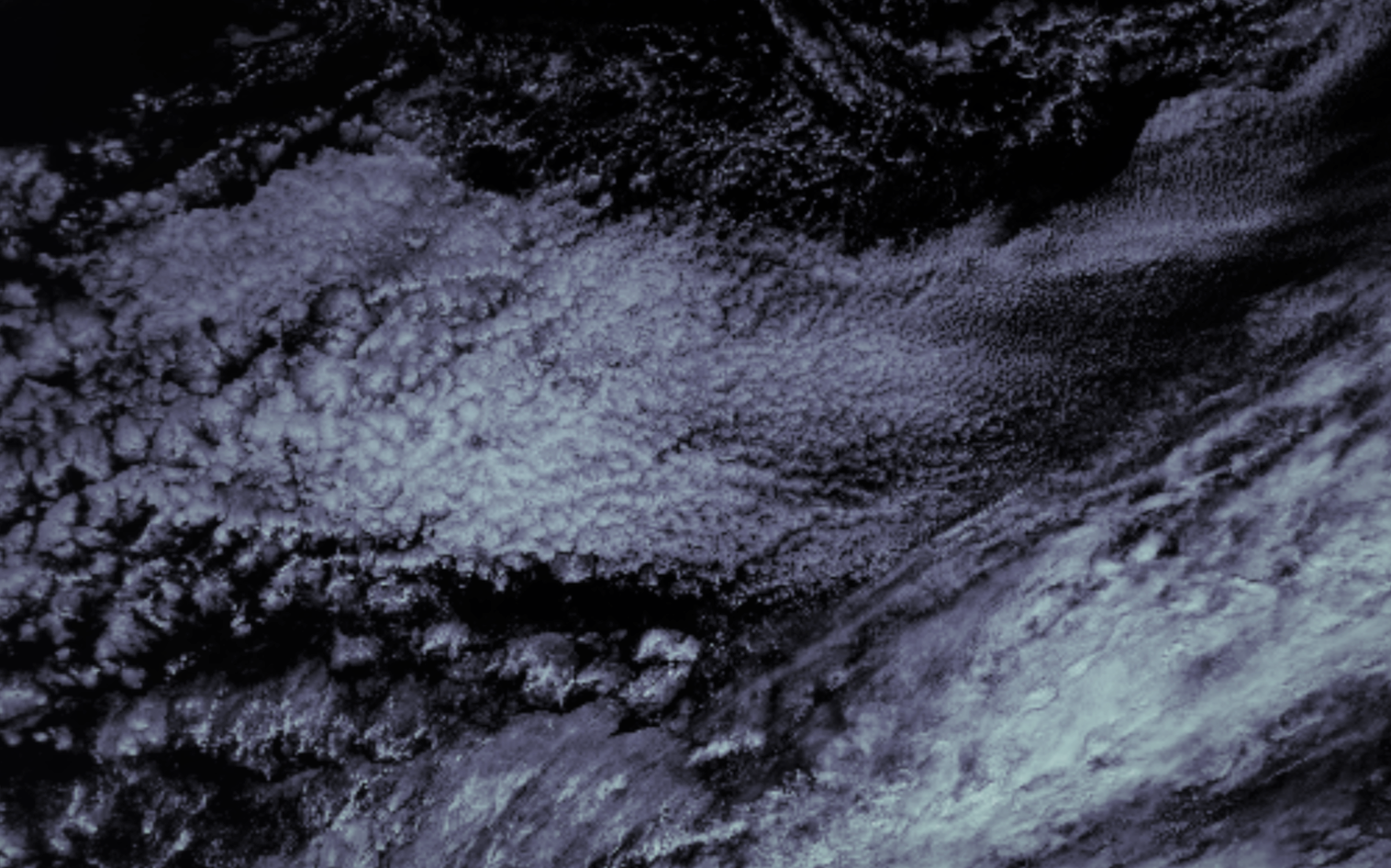}
    \end{center}
    
    \vspace{-2ex}
    
    \caption{MODIS Terra satellite visible imagery on December 1, 2015 over the East Pacific ocean, off the coast of California, capturing stratocumulus clouds at a variety of scales. Clouds at lower right are higher in altitude. We use this scene in all examples that follow.}
    %\caption{MODIS Terra satellite visible imagery on February 6, 2016 over the North East Pacific ocean, capturing hexagonal patterns of open- and closed-cell stratocumulus clouds at a variety of scales. NASA Image by Jeff Schmaltz.}
    \label{cloud.fig}
\end{figure}

Cloud classification effectively reduces the dimensionality of information in satellite images, rendering them tractable to analysis. Attempts to use neural network methods for this purpose date to the 1990s,
when Lee et al.~\cite{LeeJ} used human-labeled images to train shallow, fully connected networks to recognize stratocumulus, cumulus, and cirrus clouds. Similar efforts have continued sporadically up to the present (e.g., \cite{Bin,Wood,CloudNet}), all involving supervised learning based on images labeled with historically defined classes. However, none have produced an operational tool for automated analysis of cloud images. Supervised learning generally falls short because historical classes are artificial and are well-defined only for ``classic'' examples that make up a small fraction of total data. Furthermore, those classes do not distinguish scales of features that in nature may vary by an order of magnitude or more. For example, in the MODIS example image of Figure 1, stratocumulus clouds show a wide range of cell sizes, and cloud textures and patterns vary in complex ways.

% new computational methods can allow making use of complex information in images through unsupervised learning
Unsupervised learning methods may be a more appropriate means of making use of the complex information in large multi-spectral satellite datasets. Such methods allow
%Furthermore it is noteworthy of the limited quantities of human-labeled data, which required knowledge from domain experts as well as an extensive time commitment. Supervised learning cloud classification has the limitation for further advancement. 
%KAK: Not 100% sure if manner is the correct word here. Maybe, it should say something like "...cloud classification via supervised techniques..."
%One should consider the problem with obtaining new insight of clouds types and with scaling supervised cloud classification to such a large amount of data when compared to the more exploratory and the more time-efficient technique of unsupervised classification. 
%Historical cloud classes defined by human observers are inapppropriate for analyzing complex spatio-temporal data from multi-spectral satellite images. 
novel data-driven cloud types to emerge from imagery data, and in principle can 
%discover scientifically relevant physical features and 
track changes in cloud textures and patterns over time, by identifying both changing frequencies of individual cloud classes and evolving characteristics within a class. 

A scientifically useful operational classification system would:
\begin{enumerate}
    \item produce \textit{physically reasonable} classes with scientifically relevant distinctions
    \item capture information on cloud \textit{spatial distributions}, i.e., be not reproducible using only mean properties over the target area
    \item produce classes that in high-dimensional space are \textit{cohesive} within each class but \textit{separated} between classes  
    \item be \textit{rotationally invariant}, i.e., insensitive to the orientation of an image
    \item be \textit{stable}, i.e., produce similar or identical classes when different subsets of the data are used.
\end{enumerate}

We describe here the construction of a prototype data-driven workflow for cloud classification based on unsupervised deep learning. 
% and other machine learning methods.  (changed from "characterization:)
In the following, we introduce our model architecture and clustering procedure, apply it to images from the MODIS satellite instrument, and evaluate its ability to meet some of the key criteria listed above. 

%for scientific utility.  We evaluate our framework in two ways. First, we confirm that it classifies clear-cut cases of human-labeled data appropriately, i.e.\ that the fully unsupervised framework learns the same cloud textures classes identified by humans. Second, we evaluate whether automatically-generated classes involve commonalities of physical characteristics, i.e.\ we explore the association between the clusters and physics parameters determined by independent methods from cloud radiance data. These tests are a necessary condition for any framework to lead to scientifically useful classification schemes.

\section{Method}
%Add paragraph to summarize all method here

\subsection{Model Architecture}
We leverage recent work in self-supervised learning, in which an encoder-decoder network is trained to recover an input image. 
We use a deep convolutional autoencoder \cite{AE} to obtain dimensionally reduced information from input data. Autoencoders have been widely used to retrieve dimensionally reduced information from high-dimensional input data. The resulting lower-dimensional latent representations incorporate important input features, simplifying the classification task. In the general framework of an autoencoder, the learning process minimizes the loss function $L$:
\begin{eqnarray}\label{loss}
    \min L(x)  = \min || x - F(x) ||_{p} \hspace{2mm}, 
\end{eqnarray}
where $x$ is the input image; $F(\cdot)$ is a function which maps the input image on the dimension-reduced representation and then reconstructs the image from the intermediate information, meaning $F(x)$ is the reconstructed input image; and $|| \cdot ||_{p}$ denotes the $p$-norm of the two images. 

Our loss function combines four metrics: L1 and L2 loss, corresponding to $p$ = 1 and $p$ = 2 in Equation~\ref{loss}; the high frequency error norm after passing through the Sobel filter to detect edges of input clouds; and the multi-scale structure similarity index (MSSIM)~\cite{wang2003multiscale}, a multi-band version of SSIM~\cite{wang2004image}, an index often used in computer vision to assess image similarity. We use the Adam optimizer~\cite{adam}, a combination of RMSprop and stochastic gradient descent with momentum, to optimize our loss function, with a learning rate of 10$^{-4}$. % closely related to human eyes' perception.  

\begin{figure}
    \begin{center}
    \includegraphics[width=\columnwidth,trim=0.1in 0.1in 0.1in 0,clip]{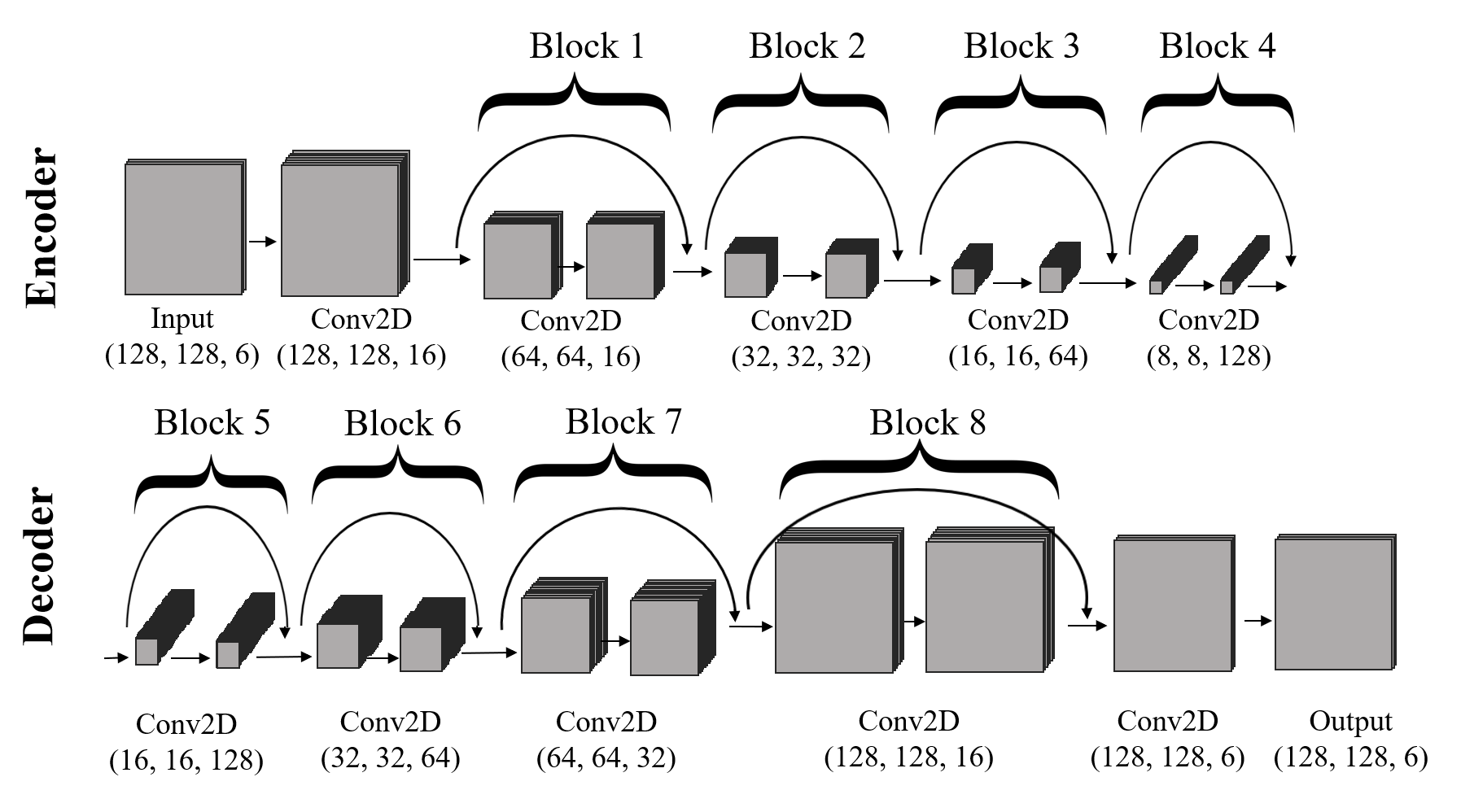}
    \end{center}
    
    \vspace{-2ex}
    
    \caption{Our autoencoder architecture. The encoder consists of four blocks, each with  two convolutional layers activated by Leaky ReLU with residual connections; 
    %We set filter size to 3$\times$3. 
    %Each block halves the image size (from 128 to 64, 32, 16, and finally 8) while increasing the number of channels (from 8 to 16, 32, 64, and finally 128). 
    the decoder has the mirror structure of the encoder.
    %and operates transposed convolution to map 1 input pixel to 3$\times$3 pixels.
    The arrows represent the flow of input images, and each bracketed triple
    gives the height, width, and channel dimensions of the layer(s) above.
    }
    \label{network.fig}
\end{figure}

We also include a convolution layer in our model in order to preserve spatial structure of the input image. The convolution operation implements a small-size filter to subset the entire image iteratively, with specified stride and width. The filter kernel operation extracts local features and parses the activation layer. The convolutional layer with activation function is described as
\begin{eqnarray}
    h^{l} &=& f \left( \sum_{i} \sum_{j} x_{(i+w-1)(j+s-1)}^{l} \otimes W^{l}_{ij} + b^{l} \right) \hspace{2mm},
\end{eqnarray}
where $h^{l}$ is the $l$th layer's latent representation; 
$f$ is a nonlinear activation function; 
$x_{(i+w)(j+s)}$ is a $w \times s$ domain for the convolutional filter; 
$W_{ij}$ is the weight at the $i$th column and $j$th row; 
$\otimes$ is the convolutional operation; 
and $b$ denotes the bias. 
We set the filter size to 3$\times$3 and use Leaky Rectified Linear Unit (Leaky ReLU) as the activation function $f(x) = \max(0.3x, x)$, as that performs better than common ReLU. 
Additionally, we build a residual connection every two convolutional layers to improve network performance, and add batch normalization after each residual connection. Between residual blocks, the size of an input image is scaled by a factor of two. In the encoder, the width and the stride are halved at each block, while the depth is doubled. In the decoder, these transformations are reversed, with the minor modification that we apply a transposed convolution kernel to map each input pixel to 3$\times$3 pixels for up-sampling. The overall model architecture is illustrated in Figure~\ref{network.fig}.

We implement the convolutional autoencoder in the TensorFlow deep learning library~\cite{abadi2016tensorflow} and use the Horovod framework~\cite{sergeev2018horovod} for data parallelization. Our encoder-decoder architecture stacks 20 convolutional layers and has \num{297232} trainable parameters, and our latent representation has size 8 $\times$ 8 $\times$ 128.  Training took \num{100000} steps and 17 hours to converge the loss function on four NVIDIA K80 GPUs on the University of Chicago's Midway compute cluster. We chose a batch size of 32 in accordance with common deep learning parameter settings and processed 789~GB of training images. We describe the input data and its pre-processing below.

% model architecture figure 

%here!!!
\subsection{Dataset}
%To examine our unsupervised approach to the cloud classification task, we construct a data alignment 
We test our workflow to perform cloud feature extraction on multi-spectral data from the MODIS instrument. Our input data is MODIS level 1B calibrated radiance imagery at 1~km resolution (MOD021KM; hereafter MOD02). This product has 36 spectral bands, from the visible to the thermal infrared; we work with bands 6, 7, 20, 28, 29, and 31, as these are the most important for the MODIS06 Level 2 algorithms that characterize cloud properties~\cite{steven17,baum12}. Bands 6, 7, and 20 (\SI{1.6}, \SI{2.1}, and \SI{3.7}{\micro\metre}) are encoded in the algorithm to estimate cloud optical properties (e.g., optical thickness and effective radius), and the brightness temperatures at bands 28, 29, and 31 (\SI{7.3}, \SI{8.5}, and \SI{11}{\micro\metre}) are used in the separation of high and low clouds and the detection of cloud phase. Note that because we seek to discover aspects of these physics variables in our classification, we do not use derived properties such as brightness temperature, but instead input radiance data directly.

%new one
To enable efficient learning of cloud features, we define the unit of our input data,  a ``patch,'' as a small subset of a typical MODIS image: 128 km $\times$ 128 km $\times$ 6 selected bands, out of an image of 2030 km $\times$ 1354 km $\times$ 36 bands. To select input patches that contain clouds, we align the MOD02 data to its corresponding MODIS35 Level 2 cloud flag product (``MOD35''), and define a patch to be valid if more than 30$\%$ of the patch is comprised of cloud pixels as detected by MOD35. We then train the network using $\sim$1.01 million patches: about 1\% of the full 19-year dataset from a single MODIS satellite instrument. 

%\subsection{Workframe}
\subsection{Clustering}
We use hierarchical agglomerative clustering (HAC) to merge data points by minimizing cluster variance, thus building a tree structure during the merging process. We choose HAC because it exhibits greater stability with respect to initialization than does k-means clustering. Our linkage metric is Ward's method, formulated as following
 \begin{eqnarray}
     \delta \text{dist}(X_A, X_B) = \frac{n_A n_B}{n_A + n_B} || C_A - C_B ||^{2} \hspace{2mm},
 \end{eqnarray}
where the distance between two clusters $X_A$ and $X_B$ is evaluated as the squared distance between the centroids of merged clusters $C_A$ and $C_B$ weighted by the number of patches in these clusters $n_A$ and $n_B$.   

% !!! Liz edits, must be checked - what is "stability"?
To choose the number of clusters for an analysis, we would ideally determine the number for which clustering results are stable (allowing the permutation of clustering categories). As an approximate measure of stability, we measure the similarity of clusters 
%in synthetic data 
using the Adjusted Mutual Information score (AMI).
%, before applying clustering to our inference data. 
We first obtain pseudo ground-truth labels by applying clustering to $\sim$320,000 patches, and then conduct tests with varying subsets of patches (chosen at random from the full dataset) and varying numbers of clusters, and compute the AMI score between the ground truth labels and subsets. %Our sanity check experiment shows that 
The AMI score typically stabilizes at cluster numbers of about 10 and higher. Most demonstration analyses shown below use 12 clusters; a larger number is likely desirable for eventual science use. %in our clustering task as cluster number 12 show reasonable correspondence with our MOD06 cloud physics parameters.

\begin{figure}[h!]
    \begin{center}
        \epsfxsize=\hsize \epsfbox{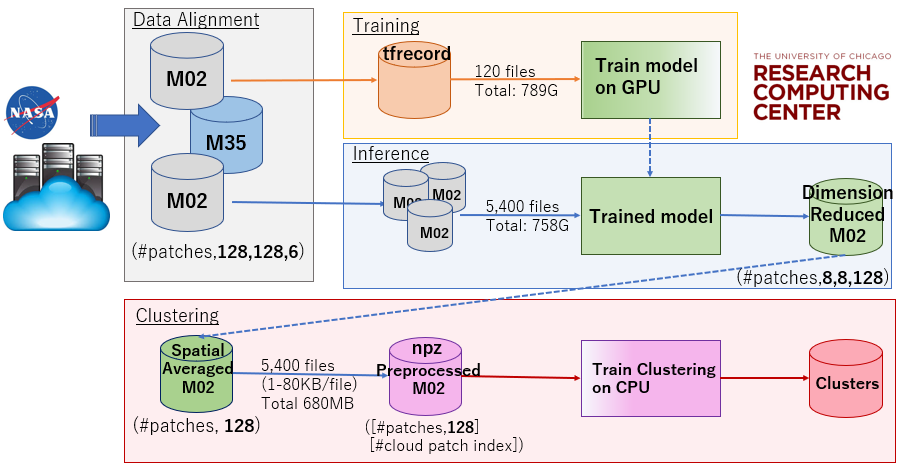}
    \end{center}
    
    \vspace{-2ex}
    
    \caption{Cartoon of the framework used in this work. M02 and M35 are the MOD02 and MOD35 satellite data products used as inputs. Orange and blue arrows show the paths taken by the training and test data, respectively; the red arrow depicts the clustering process.}
    \label{workflow.fig}
\end{figure}

Fig.~\ref{workflow.fig} shows the complete pipeline from input data to resulting clusters.

\section{Evaluation}
We report on three initial evaluations of our framework's capabilities.

\begin{figure}
    \begin{center}
    \epsfxsize=\hsize \epsfbox{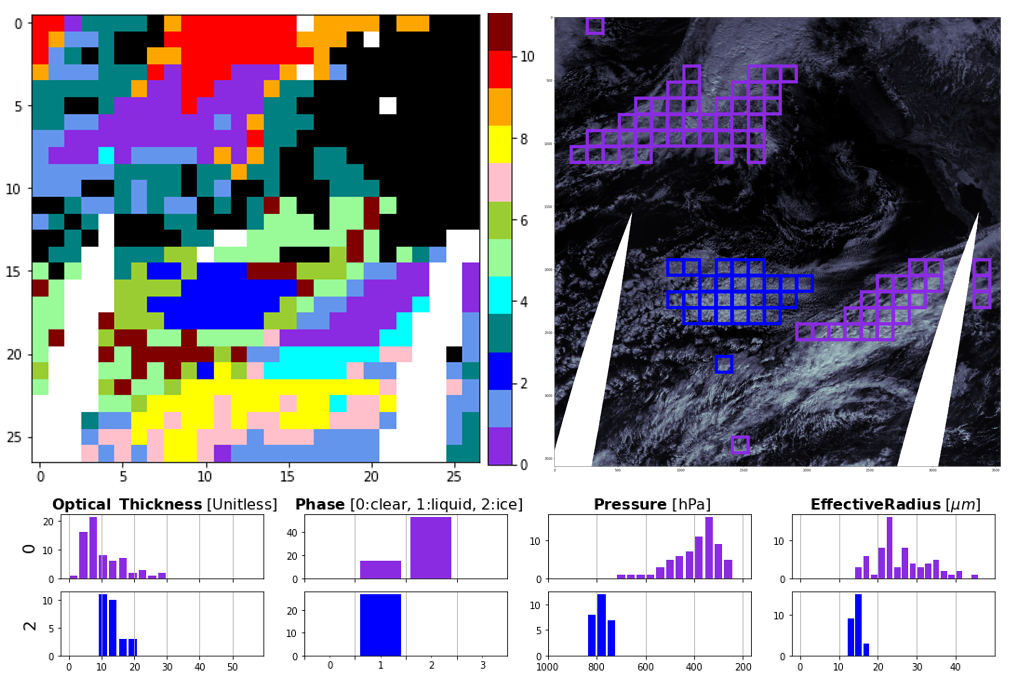}
    \end{center}
    
    \vspace{-2ex}
    
    \caption{Results of clustering results via autoencoder of MOD02 level 1B radiances in a representative MODIS swath image (December 1st, 2015, 11-44 N, 144-112 E; center of scene is shown in Fig.\ 1). Orbital coverage gaps leave missing data on sides of swath. Top left: labeled patches, classified into 12 clusters. Color bar shows cluster number in the 491 patches used; white indicates no data or invalid data; black indicates patches with  $<$30$\%$ cloud pixels. Top right: raw visible image (band 1, which is not used as input to the autoencoder), with clusters \#0 (violet) and \#2 (blue) highlighted. Bottom: histograms of path-mean values of four derived cloud physics parameters in clusters \#0 and \#2: optical thickness, phase, cloud top pressure, and effective radius. Cluster \#2 captures stratocumulus and Cluster \#0 two instances of high-altitude cirrus.}
    \label{test.fig}
\end{figure}

As a first, we evaluate the physical reasonableness of assigned cluster labels in our full workflow. That is, we ask whether clusters are associated with reasonable distributions of patch-mean values of physical variables.
Fig.~\ref{test.fig} shows results for the representative MODIS swath also shown in Figure.~\ref{cloud.fig}. Left panel shows the
 cluster labels assigned to each patch in the image; right panel shows
the raw visible image (band 1) and highlights patches assigned to two selected clusters (\#0 and \#2); and bottom panels 
 %KAK: Maybe label each image as it's own figure number for clarification instead of fig. 3 representing multiple images for clarity
show the distributions for these patches of four derived cloud physics parameters. Clustering is clearly correlated with meaningful physical cloud attributes: cluster \#2 (blue) is stratocumulus and cluster \#0 is cirrus, likely convective outflow.

%with patches characterized by low altitude (high cloud-top pressure), moderate optical thickness, and liquid-phase particles with effective radii of 10-20 microns. Cluster \#0 appears to be outflow cirrus, characterized by high altitude, low to moderate optical thickness, and predominantly ice-phase cloud particles with effective radii of 15-40 microns. 
%\#10 is the remnant of a deep convective system, characterized by high altitude, a wide range of optical thickness, and predominantly ice phase cloud particles with effective radii XX-XX microns. 

%The correlation of clustering with distinct physical characteristics is sufficient that other In other words, a fully automated unsupervised learning framework has identified high-altitude cirrus clouds, a detection that is even more impressive in the context here, where the cirrus are superimposed over lower-altitude stratocumulus.  %KAK: maybe something more like in order to connect it the model's perform: "This combination of features validates our model's performance as clouds at higher altitudes, e.g. cirrus clouds, are thin and composed of ice crystal

    \begin{figure}[h!]
    %\vspace{-2ex}
    \begin{minipage}{0.5\hsize}
        \begin{center}
        \epsfxsize=\hsize \epsfbox{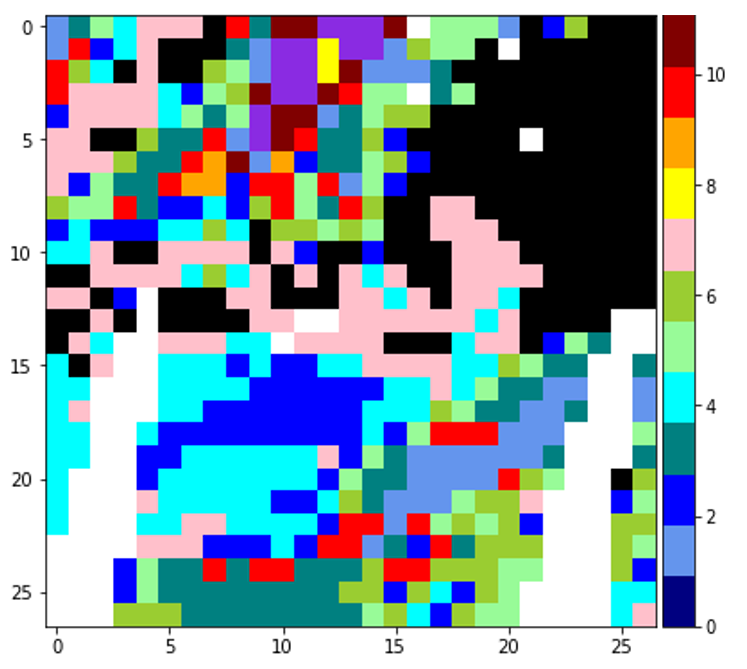}
        \end{center}
        %\vspace{-2ex}
    
    %\caption{Clustering results from patch-mean values of four MOD06 parameters, for the same image show in Figure \ref{test.fig}. Left: labeled patches, classified into 12 clusters. Color bar shows cluster number; white indicates no data or removed (invalid) data. Compare to Figure \ref{test.fig}: spatial distribution of assigned classes appears less coherent. Right: spatial distribution (heat maps) of the four MOD06 cloud physics parameters.}
    \end{minipage}
%-----------------------
    \begin{minipage}{0.485\hsize}
        \begin{center}
        \epsfxsize=\hsize \epsfbox{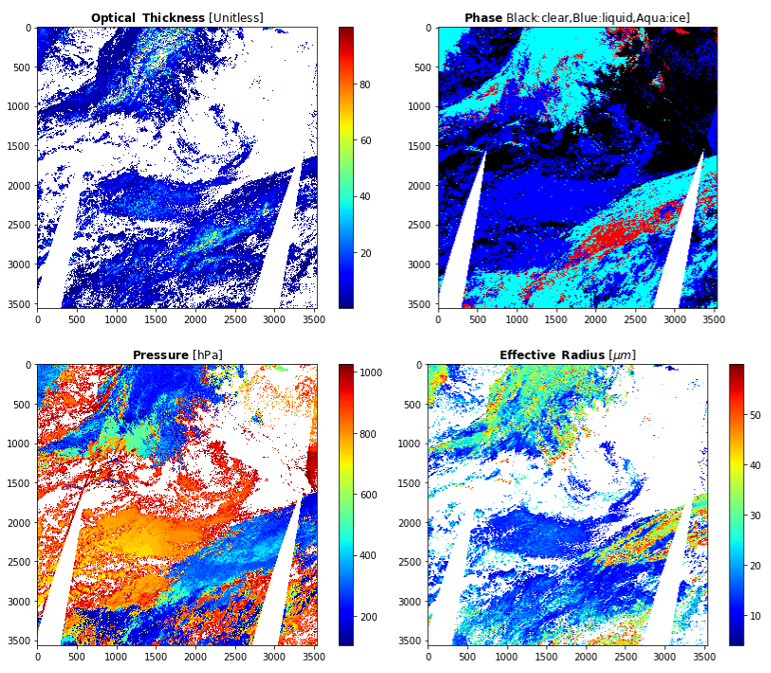}
        \end{center}

    \end{minipage}
    \vspace{-0.5ex}
    
    \caption{Results of clustering based on patch-mean values of five MOD06 parameters (the 4 shown in Fig.\ 5 plus cloud water path) in the same swath as in Fig.\ \ref{test.fig}.  Left: labeled patches, classified into 12 clusters, with same figure conventions as in Fig.\ 5. % Color bar shows cluster number; white indicates no data or invalid data; black signifies patch with $<$30$\%$ cloud pixels. 
     Right: spatial distribution (heat maps) of values of four MOD06 cloud physics parameters.
    Clustering based on patch-mean parameter value produces less spatially coherent assigned classes than in  Fig.\ \ref{test.fig} and does not capture important physical gradients, e.g.\ the sharp transition in effective radius at lower right.}
    %\label{mod06_dist.fig}
    \label{test_mod06.fig}
\end{figure}

%\vspace{-0.2in}
We then conduct a simple test of whether our clustering via autoencoder captures richer and more meaningful information on cloud distributions and properties than can be provided by the deterministic algorithms used to produce derived cloud parameters. To have scientific value, our  framework must produce information beyond that encoded in MOD06 products. We apply agglomerative clustering directly to MOD06 physics parameters and evaluate how well patches are classified without the guidance of dimension-reduced MOD02 radiance information. Results suggest that clustering via autoencoder produces classes that are spatially more cohesive and that better capture important physical transitions.
%clustering to MOD06 parameters alone cannot capture important cloud features. 
(Compare left panels of Figures~\ref{test.fig} and \ref{test_mod06.fig}.)

\begin{figure}[h!]
    \begin{center}
    \includegraphics[width=\columnwidth,trim=0.2in 0.1in 0.1in 0,clip]{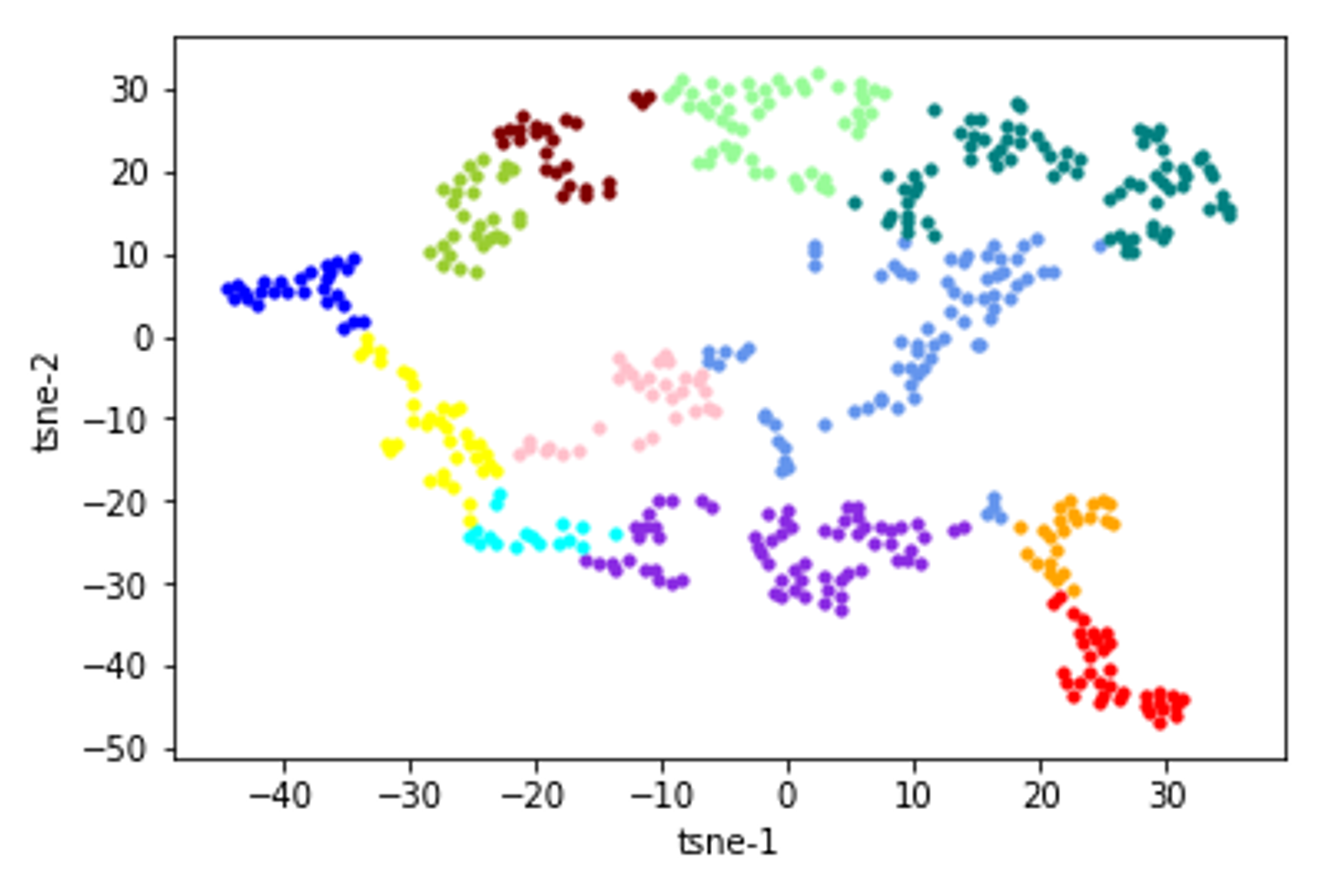}
    \end{center}
    
    \vspace{-2ex}
    \caption{A t-SNE visualization of the latent representations of
    the MOD02 patches of Figure~\ref{test.fig}, with cluster assignments represented by the same colors as in that figure.
    %tsne-1 and tsne-2 are two axis of t-SNE. 
    Patches in each cluster are projected near to each other in the t-SNE map.
    %Observe that clusters with similar physical features are projected closely in the t-SNE map.
    }
    \label{tsne.fig}
\end{figure}

 Finally, we examine the spatial distribution of the latent representation itself using t-Distributed Stochastic Neighbor Embedding (t-SNE). 
This nonlinear dimensionality reduction technique maps each point in a high-dimensional space to a two-dimensional point such that similar objects are placed near to each other and dissimilar objects far apart, with high probability~\cite{tsne08}. 
%to see whether classified patches are perceived uniformly or differently by our autoencoder. 
Resulting patch clusters  are cohesive and distinct, suggesting that 
%our latent representation successfully learns dimension-reduced unique cloud information from diverse cloud patches. Furthermore, it is demonstrated that 
agglomerative clustering within our latent representation meaningfully separates different patch types (Figure~\ref{tsne.fig}).   
%\tak{To verify whether classified data points have reasonable spacial distributions in the high dimensional space of our latent representation, we use t-SNE~\cite{tsne08} to visualize them on 2 dimensional map.

%New figure

%\begin{figure}
%    \begin{center}
%    \epsfxsize=\hsize \epsfbox{MOD06_2.PNG}
%    \end{center}
%    
%    \vspace{-2ex}
    
%    \caption{Clustering results from patch-mean values of four MOD06 parameters, for the same image show in Figure \ref{test.fig}. Left: labeled patches, classified into 12 clusters. Color bar shows cluster number; white indicates no data or removed (invalid) data. Compare to Figure \ref{test.fig}: spatial distribution of assigned classes appears less coherent. Right: spatial distribution (heat maps) of the four MOD06 cloud physics parameters.}
%    \label{test_mod06.fig}
%\end{figure}

\section{Conclusions}
% Summary of work
We describe here a prototype application of
unsupervised learning to the problem of automated classification of clouds in multi-spectral satellite imagery. 
% Summary of results
Our convolutional autoencoder generates a latent representation that, when clustered, yields physically meaningful cloud classes that pass a number of requisite tests for a scientifically useful tool. Assigned classes appropriately produce spatially coherent classifications, and capture meaningful aspects of cloud physics without being reproducible from mean values of  physics parameters alone.
%Assigned classes appropriately distinguish between commonly-accepted cloud types, produce spatially coherent classifications, and capture meaningful aspects of cloud physics without being reproducible from mean values of  physics parameters alone.%The framework produces richer information on cloud types than is available from the derived MOD06 parameters alone. 
This work supports the possibility of using unsupervised data-driven frameworks for automated cloud classification and pattern discovery without requiring the prior hypothesis of ground-truth labeled data. 
While results here are preliminary, they suggest that similar frameworks can be used to analyze multi-year, global data to track short-term evolution during cloud lifecycles and long-term trends in the distribution of cloud features and characteristics. 

More generally,
unsupervised learning methods have broad potential applicability in the Earth sciences. In the satellite era, a primary challenge to environmental scientists is not gathering  data but finding meaning in overwhelmingly large datasets. Unsupervised learning has the potential to reveal patterns directly learned from observation, which can then provide new insights and help diagnose drivers of system behavior.
%, and suggest fruitful new research directions. 

\section*{Acknowledgments}
This work was supported by the % performed as part of the 
%U.\ Chicago 
Center for Robust Decision-making on Climate and Energy Policy (RDCEP), %supported by
%under the NSF Decision Making Under Uncertainty Program, 
NSF
award SES-1463644, and 
%was performed on 
used
computers at the U.Chicago Research Computing Center and the Argonne Leadership Computing Facility, a DOE Office of Science User Facility, contract DE-AC02-06CH11357.

\bibliographystyle{ieeetr}
\bibliography{ci_references}
\end{document}